\documentstyle[graphicx,12pt]{article}
\begin{document}

\small
\hoffset=-1truecm
\voffset=-2truecm
\title{\bf The Casimir effect for parallel plates involving massless Majorana fermions at
finite temperature}
\author{Hongbo Cheng\footnote {E-mail address:
hbcheng@sh163.net}\\
Department of Physics, East China University of Science and
Technology,\\ Shanghai 200237, China\\
The Shanghai Key Laboratory of Astrophysics,\\ Shanghai 200234,
China}

\date{}
\maketitle

\begin{abstract}
We study the Casimir effect for parallel plates with massless
Majorana fermions obeying the bag boundary conditions at finite
temperature. The thermal influence will modify the effect. It is
found that the sign of the Casimir energy keeps negative if the
product of plate distance and the temperature is larger than a
special value or the energy will change to be positive. The
Casimir energy rises with the stronger thermal influence. We show
that the attractive Casimir force between two parallel plates
becomes greater with the increasing temperature. In the case of
piston system involving the same Majorana fermions with the same
boundary conditions, the attractive force on the piston will
weaker in the hotter surrounding.
\end{abstract}
\vspace{6cm} \hspace{1cm} PACS number(s): 03.70.+k, 03.65.Ge

\newpage

\noindent \textbf{I.\hspace{0.4cm}Introduction}

The Casimir effect is essentially a direct consequence of quantum
field theory subject to a change in the spectrum of vacuum
oscillations when the quantization volume is bounded, or some
background field is inserted. Historically H. B. G. Casimir
originated the effect of boundaries [1]. Twenty years later T. H.
Boyer found that the Casimir force for a conducting spherical
shell is repulsive, showing that the circumstances will determine
the force nature [2]. Afterwards more efforts have been paid for
the problem and related topics and more and more results and
methods have been put forward [3-10]. The precision of the
measurements has been greatly improved expentaly. Recently the
sign of Casimir energy and the nature of Casimir force have been
applied in many subjects because the Casimir effect depends on
various factors. For example the Casimir effect can be utilized to
study the high-dimensional spacetimes. We can research on the
Casimir effect for parallel plates in the spacetimes with extra
compactified dimensions to show that the extra-dimension influence
is manifest and distinct while the model was introduced [11-21].
The research on the Casimir effect for parallel plates or piston
in the braneworld such as Randall-Sundrum models and etc. have
also made great progress [22-31]. It is interesting that we can
estimate the properties of the world with more than four
dimensions. The Casimir effect has been also explored in the
context of string theory [32].

It is significant to study the Casimir effect for Majorana
fermions under some kinds of boundary conditions. Majorana
fermions have been discussed widely in many areas of physics. As a
Majorana fermion the Kaluza-Klein neutrino could be thought as a
candidate of weakly interacting massive particle [33]. In the
physics of superconductors and superfluids a Majorana bound state
is theoretically predicted in rotating superfluid $^{3}He-A$
between parallel plates with the gap that is about $10\mu m$ and
in triplet superconductors [34]. The Majorana vortex bound state
is subject to a singular vortex in chiral p-wave superfluid [35].
Further the Casimir effect for Majorana fermions in parallel
plates with bag or chiral boundary conditions is studied [36]. It
was pointed that the Casimir energy is singularity free. In
addition the Casimir effect for a massive fermionic field in the
geometry of two parallel plates in the Minkowski spacetime with an
arbitrary number of toroidally compactified spatial dimensions was
also evaluated [37].

The quantum field theory at finite temperature shares a lot of
effects. The thermal influence on the Casimir effect can not be
neglected and the influence certainly modifies the effect. The
Casimir energy for a rectangular cavity under a nonzero
temperature environment was considered and the temperature
controls the energy sign [38]. The Casimir effect for parallel
plates including thermal corrections in the spacetime with
additional compactified dimensions was explored and the magnitude
of Casimir force as well as the sign of Casimir energy change with
the temperature [39-42]. In addition the Casimir effect for a
scalar field within two parallel plates under thermal influencd in
the bulk region of Randall-Sundrum models was also evaluated [43].

In this paper we are going to consider the thermal corrections to
the Casimir effect for parallel plates as well as pistons
containing massless Majorana fermions in detail to generalize some
results of Ref. [36]. The description of the Casimir effect for
massless Majorana fermions in the confine region such as parallel
plates will change at different temperature. At first we derive
the frequency of massless Majorana fermions subject to bag
boundary conditions with thermal corrections by means of
finite-temperature field theories. We regularize the frequency to
obtain the Casimir energy density with the help of zeta function
technique. Further the Casimir force between the parallel plates
involving the Majorana fermions can also be gained from the
Casimir energy density at finite temperature. Certainly the
discussions can be generalized to the case of piston. It is
necessary to explore the nature of Casimir energy and Casimir
force for massless Majorana fermions within the parallel-plate
system. In the end the conclusions are listed.

\vspace{3cm}

\noindent \textbf{II.\hspace{0.4cm}The Casimir effect of massless
Majorana fermions in the parallel-plate system with bag boundary
conditions at finite temperature}

In finite-temperature field theories the imaginary time formalism
can be utilized to describe the fermion field in the thermal
equilibrium. We introduce a partition function for a system
containing Dirac fields,

\begin{equation}
Z=N\int_{antiperiodic}D\overline{\psi}D\psi\exp[\int_{0}^{\beta}
\int d^{3}x\mathcal{L}(\psi, \partial_{E}\psi)]
\end{equation}

\noindent where $\mathcal{L}$ is the Lagrangian density for the
system under consideration. $N$ a constant and "antiperiodic"
means,

\begin{equation}
\psi(0, \mathbf{x})=-\psi(\tau=\beta, \mathbf{x})
\end{equation}

\noindent where $\beta=\frac{1}{T}$ is the inverse of the
temperature and $\tau=it$. Here the massless Majorana fermions can
be described through the Dirac equation as follow,

\begin{equation}
i\gamma^{\mu}\partial_{\mu}\psi=0
\end{equation}

\noindent where $\mu=0, 1, 2, 3$. These massless fields satisfying
the MIT bag boundary conditions which are expressed as,

\begin{equation}
i n^{\mu}\gamma_{\mu}\psi=\psi
\end{equation}

\noindent are confined to the interior of parallel-plate system.
In Eq. (4) $n^{\mu}=(0, \mathbf{n})$ and $\mathbf{n}$ the vectors
normal to the surface of plates and directed to the interior of
the slab configuration. According to the solutions to the
equations of motion (3) and the boundary conditions (4), the
generalized zeta function can be written as,

\begin{eqnarray}
\zeta(s, -\partial_{E})=Tr(-\partial_{E})^{-s}\nonumber\hspace{5cm}\\
=-\int\frac{d^{2}k}{(2\pi)^{2}}\sum_{n=0}^{\infty}\sum_{l=-\infty}^{\infty}
[k^{2}+\frac{\pi^{2}(n+\frac{1}{2})^{2}}{R^{2}}+\frac{4\pi^{2}(l+\frac{1}{2})^{2}}
{\beta^{2}}]^{-s}
\end{eqnarray}

\noindent where

\begin{equation}
\partial_{E}=\frac{\partial^{2}}{\partial\tau^{2}}+\nabla^{2}
\end{equation}

\noindent and $k$ denote the transverse components of the
momentum. $R$ is the distance of the plates. We obtain the total
energy density of the system with thermal corrections in virtue of
$\varepsilon=-\frac{\partial}{\partial\beta}(\frac{\partial\zeta(s;
-\partial_{E})}{\partial s}|_{s=0})$ and regularize the expression
by means of zeta function technique to obtain the Casimir energy
per unit area for parallel plates with massless Majorana fermions
at finite temperature as follow,

\begin{eqnarray}
\varepsilon_{C}=-\frac{7}{8}\frac{\pi^{2}}{720R^{3}}+\frac{1}{\sqrt{2}}
\frac{1}{\beta^{\frac{3}{2}}R^{\frac{3}{2}}}\sum_{n_{1}=1}^{\infty}
\sum_{n_{2}=0}^{\infty}n_{1}^{-\frac{3}{2}}(n_{2}+\frac{1}{2})^{\frac{3}{2}}
K_{\frac{3}{2}}[\frac{\pi\beta}{R}n_{1}(n_{2}+\frac{1}{2})]\nonumber\\
+\frac{\pi}{\sqrt{2}}\frac{1}{\beta^{\frac{1}{2}}R^{\frac{5}{2}}}
\sum_{n_{1}=1}^{\infty}\sum_{n_{2}=0}^{\infty}n_{1}^{-\frac{1}{2}}
(n_{2}+\frac{1}{2})^{\frac{5}{2}}\nonumber\hspace{5cm}\\\times
[K_{\frac{1}{2}}(\frac{\pi\beta}{R}n_{1}(n_{2}+\frac{1}{2}))
+K_{\frac{5}{2}}(\frac{\pi\beta}{R}n_{1}(n_{2}+\frac{1}{2}))]\hspace{2cm}
\end{eqnarray}

\noindent where $K_{\nu}(z)$ is the modified Bessel function of
the second kind. Here we hire the zeta function regularization. In
Eq. (7) the terms with series converge very quickly and only the
first several summands need to be taken into account for numerical
calculation to further discussion. If the temperature approaches
zero, the Casimir energy density will turn to be the results of
Ref. [36], so will the Casimir force. When the temperature becomes
extremely high the asymptotic behaviour of the Casimir energy per
unit area for massless Majorana fermions within the parallel
plates is,

\begin{equation}
\lim_{T\longrightarrow\infty}\varepsilon_{C}=\frac{7}{240}
\frac{\pi}{\Gamma(\frac{3}{2})}\frac{R}{\beta^{4}}
\end{equation}

\noindent meaning that the Casimir energy rises as the temperature
grows. Having performed the burden calculation, we find that the
sign of the Casimir energy will be positive if the plates
separation and temperature satisfy as $RT>0.37$ in natural unit.
The special value has something to do with the systems and
boundary conditions.

The Casimir force on the plates is given by the derivative of the
Casimir energy with respect to the plate distance. Here the
Casimir force per unit area on the plates enclosing the massless
Majorana fermions with bag boundary conditions can be written as,

\begin{eqnarray}
f_{C}=-\frac{\partial\varepsilon_{C}}{\partial R}\nonumber\hspace{9.5cm}\\
=-\frac{7\pi^{2}}{1920}\frac{1}{R^{4}}+\frac{3}{2^{\frac{3}{2}}}
\frac{1}{\beta^{\frac{3}{2}}R^{\frac{5}{2}}}\sum_{n_{1}=1}^{\infty}
\sum_{n_{2}=0}^{\infty}n_{1}^{-\frac{3}{2}}(n_{2}+\frac{1}{2})^{\frac{3}{2}}
K_{\frac{3}{2}}[\frac{\pi\beta}{R}n_{1}(n_{2}+\frac{1}{2})]\nonumber\\
-\frac{\pi}{2^{\frac{3}{2}}}\frac{1}{\beta^{\frac{1}{2}}R^{\frac{7}{2}}}
\sum_{n_{1}=1}^{\infty}\sum_{n_{2}=0}^{\infty}n_{1}^{-\frac{1}{2}}
(n_{2}+\frac{1}{2})^{\frac{5}{2}}\nonumber\hspace{4.5cm}\\
\times[K_{\frac{1}{2}}(\frac{\pi\beta}{R}n_{1}(n_{2}+\frac{1}{2}))
+K_{\frac{5}{2}}(\frac{\pi\beta}{R}n_{1}(n_{2}+\frac{1}{2}))]\nonumber\\
+\frac{5}{2^{\frac{3}{2}}}\frac{1}{\beta^{\frac{1}{2}}R^{\frac{7}{2}}}
\sum_{n_{1}=1}^{\infty}\sum_{n_{2}=0}^{\infty}n_{1}^{-\frac{1}{2}}
(n_{2}+\frac{1}{2})^{\frac{5}{2}}\nonumber\hspace{4.5cm}\\
\times[K_{\frac{1}{2}}(\frac{\pi\beta}{R}n_{1}(n_{2}+\frac{1}{2}))
+K_{\frac{5}{2}}(\frac{\pi\beta}{R}n_{1}(n_{2}+\frac{1}{2}))]\nonumber\\
-\frac{\pi^{2}}{2^{\frac{3}{2}}}\frac{\beta^{\frac{1}{2}}}{R^{\frac{9}{2}}}
\sum_{n_{1}=1}^{\infty}\sum_{n_{2}=0}^{\infty}n_{1}^{\frac{1}{2}}
(n_{2}+\frac{1}{2})^{\frac{7}{2}}\nonumber\hspace{5cm}\\
\times[K_{\frac{1}{2}}(\frac{\pi\beta}{R}n_{1}(n_{2}+\frac{1}{2}))
+2K_{\frac{3}{2}}(\frac{\pi\beta}{R}n_{1}(n_{2}+\frac{1}{2}))\nonumber\\
+K_{\frac{7}{2}}(\frac{\pi\beta}{R}n_{1}(n_{2}+\frac{1}{2}))]\hspace{2cm}
\end{eqnarray}

\noindent Similarly if the temperature vanishes, the Casimir
pressure shown in Eq. (9) keeps the first term that is consistent
with that from Ref. [36]. When the two plates leave apart from
each other, the Casimir force per unit area does not vanish
because of the temperature and is expressed as,

\begin{equation}
\lim_{R\longrightarrow\infty}f_{C}=-\frac{7\pi^{2}}{120}T^{4}
\end{equation}

\noindent It is evident that the Casimir force between the two
plates where the Majorana fermions satisfy the bag boundary
conditions becomes greater during the process that the temperature
grows while the force nature remains attractive. As a function of
plate distance for a definite temperature the Casimir force per
unit area with thermal influence is plotted in Fig. 1. The curves
of Casimir pressure associated with the plates separation for
different magnitude of temperature are similar. As the temperature
becomes higher the whole Casimir pressure curve will lower, which
means that the stronger thermal influence leads the attractive
Casimir force for two parallel plates containing massless Majorana
fermions to be greater. It is interesting that here the asymptotic
value of the Casimir force with thermal corrections does not
vanish unless the temperature is equal to zero. It should also be
pointed out that the asymptotic behaviour of the Casimir force
shown in Eq. (10) is independent of plates position.

\vspace{3cm}

\noindent \textbf{III.\hspace{0.4cm}The Casimir piston of massless
Majorana fermions with bag boundary conditions at finite
temperature}

It is significant to explore the Casimir effect for massless
Majorana within the device of piston at finite temperature. More
efforts have been contributed to the topics related to the piston
[15-22, 36]. Under the environment with nonzero temperature we
discuss the massless Majorana fermions in the system consisting of
three parallel plates and the bag boundary conditions are imposed
on the plates. Having found and regularized the total vacuum
energy per unit area of the three-parallel-plate system, we obtain
the Casimir energy per unit area
$\varepsilon_{C}=\varepsilon_{C}^{1}(R,
T)+\varepsilon_{C}^{2}(L-R, T)+\varepsilon_{C}^{out}(T)$, where
$\varepsilon_{C}^{1}(R, T)$, $\varepsilon_{C}^{2}(L-R, T)$ and
$\varepsilon_{C}^{out}(T)$ represent the Casimir energy per unit
area of two inner parts and the Casimir energy per unit area
outside the system respectively. Here $\varepsilon_{C}^{1}(R,
T)=\varepsilon_{C}$ presented in Eq. (7) and
$\varepsilon_{C}^{2}(L-R, T)=\varepsilon_{C}|_{R\longrightarrow
L-R}$. The Casimir energy per unit area outside the
three-parallel-plate system has nothing to do with the plates
distance. Further the Casimir force per unit area on the piston is
given with the help of derivative of the Casimir energy per unit
area with respect to the plates separation like
$f'_{PC}=-\frac{\partial\varepsilon_{C}}{\partial
R}=-\frac{\partial}{\partial R}\varepsilon_{C}^{1}(R,
T)-\frac{\partial}{\partial R}\varepsilon_{C}^{2}(L-R,
T)-\frac{\partial}{\partial R}\varepsilon_{C}^{out}(T)$. As one
outer plate is moved to the extremely distant place or
equivalently $L\longrightarrow\infty$, the Casimir pressure
between the remaining two plates with Majorana fermions is,

\begin{eqnarray}
f_{PC}=\lim_{L\longrightarrow\infty}f'_{PC}\nonumber\hspace{7.5cm}\\
=\frac{3}{\sqrt{2}}\frac{1}{R^{\frac{3}{2}}\beta^{\frac{5}{2}}}
\sum_{n_{1}=1}^{\infty}\sum_{n_{2}=0}^{\infty}n_{1}^{-\frac{3}{2}}
(n_{2}+\frac{1}{2})^{\frac{3}{2}}K_{\frac{3}{2}}(\frac{4\pi
R}{\beta}n_{1}(n_{2}+\frac{1}{2}))\nonumber\hspace{1cm}\\
+\frac{(6+2\pi)\sqrt{2}}{R^{\frac{1}{2}}\beta^{\frac{7}{2}}}
\sum_{n_{1}=1}^{\infty}\sum_{n_{2}=0}^{\infty}n_{1}^{-\frac{1}{2}}
(n_{2}+\frac{1}{2})^{\frac{5}{2}}[K_{\frac{1}{2}}(\frac{4\pi
R}{\beta}n_{1}(n_{2}+\frac{1}{2}))\nonumber\\
+K_{\frac{5}{2}}(\frac{4\pi
R}{\beta}n_{1}(n_{2}+\frac{1}{2}))]\nonumber\hspace{1cm}\\
-2^{\frac{7}{2}}\pi^{2}\frac{R^{\frac{1}{2}}}{\beta^{\frac{9}{2}}}
\sum_{n_{1}=1}^{\infty}\sum_{n_{2}=0}^{\infty}n_{1}^{\frac{1}{2}}
(n_{2}+\frac{1}{2})^{\frac{7}{2}}[K_{\frac{1}{2}}(\frac{4\pi
R}{\beta}n_{1}(n_{2}+\frac{1}{2}))\nonumber\hspace{0.5cm}\\
+2K_{\frac{3}{2}}(\frac{4\pi R}{\beta}n_{1}(n_{2}+\frac{1}{2}))
+K_{\frac{7}{2}}(\frac{4\pi R}{\beta}n_{1}(n_{2}+\frac{1}{2}))]
\end{eqnarray}

\noindent In fact the piston device can help us to cancel the
terms which are independent of the plate gap. According to Eq.
(11), we discover that,

\begin{equation}
\lim_{R\longrightarrow\infty}f_{PC}=0
\end{equation}

\noindent which is favoured by the observational results and,

\begin{equation}
\lim_{T\longrightarrow\infty}f_{PC}=0
\end{equation}

\noindent which means that the Casimir force is weaker with
stronger thermal influence. The dependence of the Casimir force
per unit area for the piston on the distance between the piston
and the other plate is shown in Fig. 2. It is manifest that the
attractive Casimir force for Majorana fermion piston will be
weaker as the temperature grows because the parts in the Casimir
pressure like Eq. (10) are plate-position-independent and are
certainly cancelled in the piston device although their magnitude
is an increasing function of temperature. The curves of the force
with different temperature are similar.

\vspace{3cm}

\noindent \textbf{IV.\hspace{0.4cm}Conclusion}

In this work we investigate the Casimir effect with thermal
modification for parallel plates and piston filled with massless
Majorana fermions while there are bag boundary conditions imposed
at the plates. At first we discover that the Casimir energy for
two parallel plates rises with the increase in the temperature.
When the product of the plates separation and the temperature is
larger than a special value like $RT>0.37$, the sign of the
Casimir energy will change to be positive. The special value will
be different in different system with different boundary
conditions. It is also found that the magnitude of Casimir force
between two parallel plates with massless Majorana fermions
satisfying the bag boundary conditions at the plates become
greater as the thermal influence is stronger while the Casimir
force keeps attractive. In the case of piston system which
contains the massless Majorana fermions with the same bag boundary
conditions at the plates, the attractive Casimir force on the
piston will be weaker when the surrounding is hotter. As the
temperature is extremely high, the Casimir force on the piston
will vanish. When the plates distance is extremely large, the
asymptotic value of the Casimir force between two parallel plates
depends on the temperature and does not disappear. In the context
of piston model the Casimir force between the piston and the other
plate will approach to be zero when the two plates leave each
other much farther. Our results can return to be those of Ref.
[36] when the temperature is equal to zero. Here we declare how
the Casimir effect for massless Majorana fermions within the
parallel system subject to the bag boundary conditions depends on
the temperature.

\vspace{3cm}

\noindent\textbf{Acknowledgement}

This work is supported by NSFC No. 10875043 and is partly
supported by the Shanghai Research Foundation No. 07dz22020.

\newpage
\begin{figure}
\setlength{\belowcaptionskip}{10pt} \centering
  \includegraphics[width=15cm]{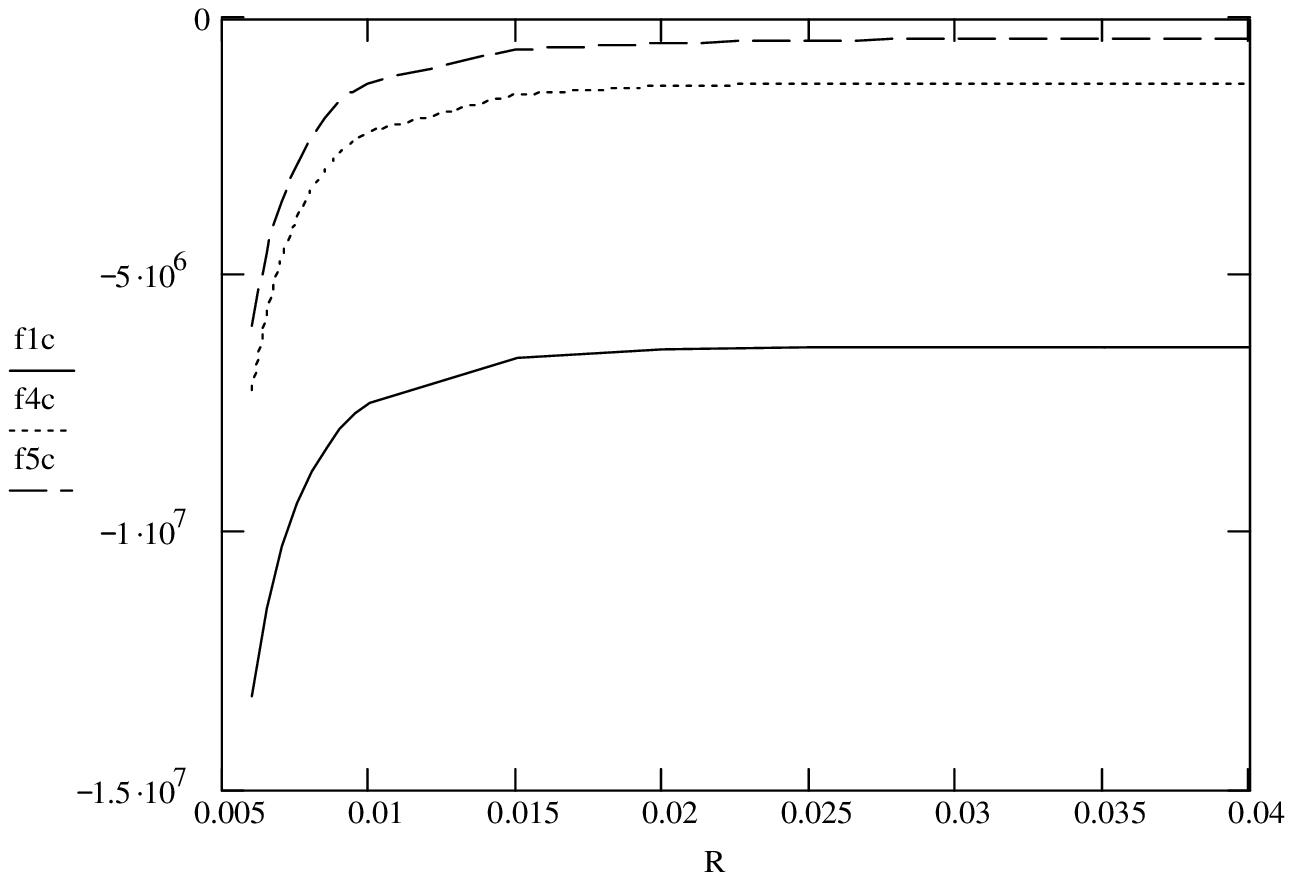}
  \caption{The solid, dot and dashed curves of the Casimir force per
  unit area between two parallel plates as functions of plate separation
  $R$ for $\beta=0.02, 0.03, 0.04$ respectively.}
\end{figure}

\newpage
\begin{figure}
\setlength{\belowcaptionskip}{10pt} \centering
  \includegraphics[width=15cm]{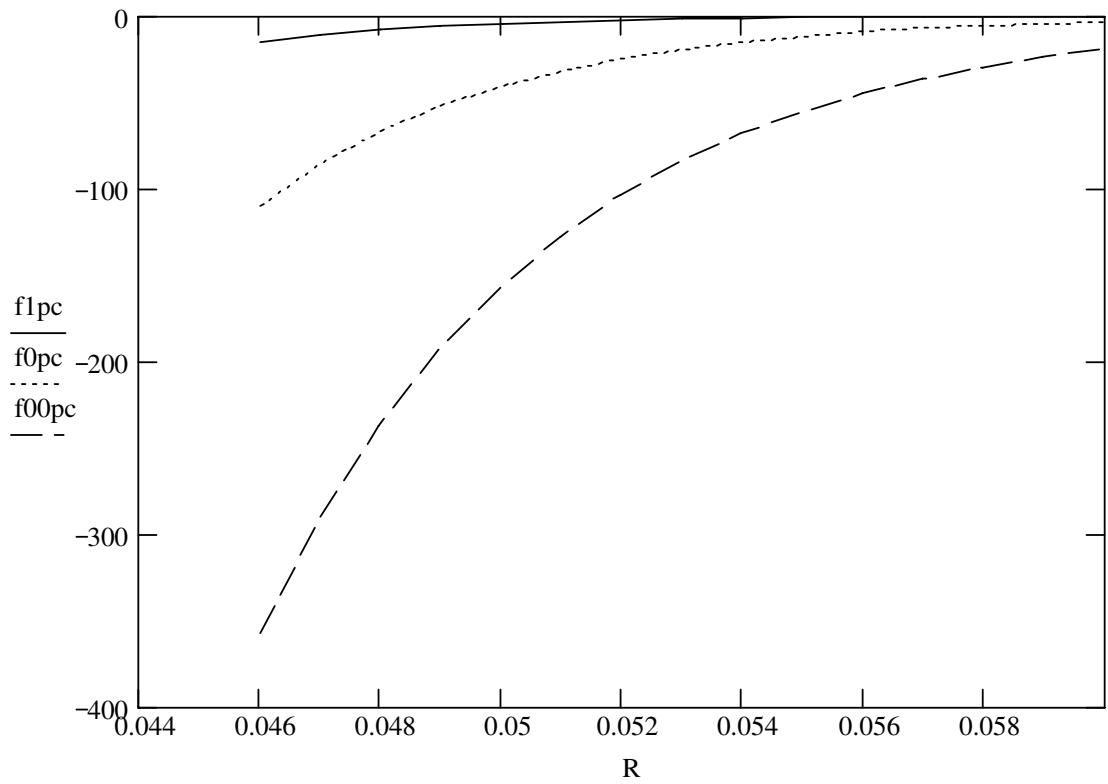}
  \caption{The solid, dot and dashed curves of the Casimir force per
  unit area for piston as functions of plate separation
  $R$ for $\beta=0.02, 0.025, 0.03$ respectively.}
\end{figure}

\end{document}